\documentclass[amstex,twocolumn,showpacs,floats,floatfix,superscriptaddress,aps,pra]{revtex4}
\usepackage{amsfonts}
\usepackage{amssymb}
\usepackage{amsmath}
\usepackage{calc}
\usepackage{graphicx}
\usepackage{bm}

\def\be{ \begin{equation} }
\def\ee{ \end{equation} }

\def\LZphase{\phi}

\def\EVA{\Lambda}

\def\Omeg0{\Omega_0}

\def\Pi{P_{1 \rightarrow 3}}

\def\P22{P_{2 \rightarrow 2}}

\begin{document}

\author{A. A. Rangelov}
\affiliation{Department of Physics, Sofia University, James Bourchier 5 blvd, 1164 Sofia, Bulgaria}
\author{J. Piilo}
\affiliation{Department of Physics, Sofia University, James Bourchier 5 blvd, 1164 Sofia, Bulgaria}
\affiliation{School of Pure and Applied Physics, University of KwaZulu-Natal, Durban 4041, South Africa}
\author{N. V. Vitanov}
\affiliation{Department of Physics, Sofia University, James Bourchier 5 blvd, 1164 Sofia, Bulgaria}
\affiliation{Institute of Solid State Physics, Bulgarian Academy of Sciences,
 Tsarigradsko chauss\'{e}e 72, 1784 Sofia, Bulgaria}
\title{Counterintuitive transitions between crossing energy levels}
\date{\today }

\begin{abstract}
We calculate analytically the probabilities for intuitive and counterintuitive transitions in a three-state system,
 in which two parallel energies are crossed by a third, tilted energy. 
The state with the tilted energy is coupled to the other two states in a chainwise linkage pattern
 with constant couplings of finite duration. 
The probability for a counterintuitive transition is found to increase with the square of the coupling and
 decrease with the squares of the interaction duration, the energy splitting
 between the parallel energies and the tilt (chirp) rate. 
Physical examples of this model can be found in coherent atomic excitation and optical shielding in cold atomic collisions.
\end{abstract}

\maketitle


\section{Introduction}

The famous Landau-Zener (LZ) model \cite{LZ} is the most popular tool for
estimating the transition probability between two states whose energies
cross in time. This model assumes a constant interaction of infinite
duration and linear energies. Owing to some mathematical subtleties, the LZ
model often provides more accurate results than anticipated (given the
rather simple time dependence of the LZ Hamiltonian) when applied to real
physical systems with more sophisticated time dependences. The popularity of
the LZ model is further motivated by the extreme simplicity of the
transition probability.

The LZ model has been extended to three and more levels by a number of
authors. There are two main types of generalizations: single-crossing
(bow-tie) models and multiple-crossings grid models.

In the bow-tie models all state energies cross at the same instant of time.
Carroll and Hioe have solved the three-state bow-tie model in a special
symmetric case \cite{Carroll-special} and in the general case \cite%
{Carroll-general}. An extension of this model to $N$ states has been
suggested \cite{Harmin,Brundobler} and then rigorously derived by Ostrovsky
and Nakamura \cite{Ostrovsky97}. A further extension, wherein one of the
levels is split into two parallel levels, has been suggested \cite{Demkov00}
and derived \cite{Demkov01} by Demkov and Ostrovsky. An example of a bow-tie
system occurs when a sequentially coupled quantum ladder of states in an atom
or a molecule is driven by chirped laser pulses \cite{ARPC}; an adiabatic
sweep of frequency through resonance will transfer all population from the
lowest to the highest energy state, for either sign of the chirp. A bow-tie
type of linkage can also occur in a rf-pulse controlled Bose-Einstein
condensate output coupler \cite{Mewes,VitanovBEC}. Yet another example is
the coupling pattern of Rydberg sublevels in a magnetic field \cite{Harmin}.

In the multiple-crossings grid models the energy diagram consists of a grid of
crossings formed by two manifolds of rectilinear parallel diabatic energies that
cross each other. In the Demkov-Osherov (DO) model \cite{Demkov68,Kayanuma},
a single tilted energy crosses a set of $N$ parallel energies. This model
has been generalized to the case when the single tilted energy is replaced
by a set of $M$ parallel energies, which cross the other set of $N$
parallel energies \cite{Demkov95,Usuki,Ostrovsky98}. The special case when $M
$ and $N$ are infinite (so that the grid of crossings is periodic) has also
been solved \cite{Demkov95b}. Effects of level degeneracies and
quasi-degeneracies have been studied by Yurovsky and Ben-Reuven \cite{Yurovsky98,Yurovsky01}.

In the most general case of an asymmetric linear Hamiltonian, $\mathsf{H}(t)=%
\mathsf{A}+\mathsf{B}t$, where $\mathsf{B}$ is diagonal, the general
solution has not been derived yet, but exact results for some
\textquotedblleft survival\textquotedblright\ probabilities have been
conjectured \cite{Brundobler} and derived \cite%
{Shytov,Sinitsyn04,Volkov04,Volkov05}.

A variety of physical systems provide examples of multiple level crossings.
Amongst them we mention
 ladder climbing of atomic and molecular states by chirped laser pulses \cite{climbing},
 optical shielding in cold atomic collisions \cite{Suominen96a,Suominen96b,Yurovsky97,Napolitano,Piilo},
 optical centrifuge for molecules \cite{centrifuge},
 Stark-chirped rapid adiabatic passage (SCRAP) \cite{SCRAP-theory,SCRAP-experiment,SCRAP-3S},
 and creation of entaglement in many-particle systems \cite{entanglement}.
In fact, physical situations where multiple level crossings play a role
 have been discussed already in 1960's, when the harpoon model for reactive scattering was considered \cite{Child}. 
If one adds the vibrational states to the scattering picture, then one obtains a multilevel crossing model,
 which has been studied using the LZ model \cite{Child}.

A general feature of all soluble multilevel models is that the transition
probabilities $P_{m\rightarrow n}$ between states $\psi _{m}$ and $\psi _{n}$
are given by very simple expressions, as in the original LZ model, although
the derivations are usually quite involved. In the grid models in
particular, such as the DO model, the exact probabilities $P_{m\rightarrow n}
$ have the same form --- products of LZ probabilities for transition or
no-transition applied at the relevant crossings --- as what would be
obtained by naive multiplication of LZ probabilities while moving across the
grid of crossings from $\psi _{m}$ to $\psi _{n}$, without accounting for phases and interferences.

\begin{figure}[t]
\includegraphics[width=75mm]{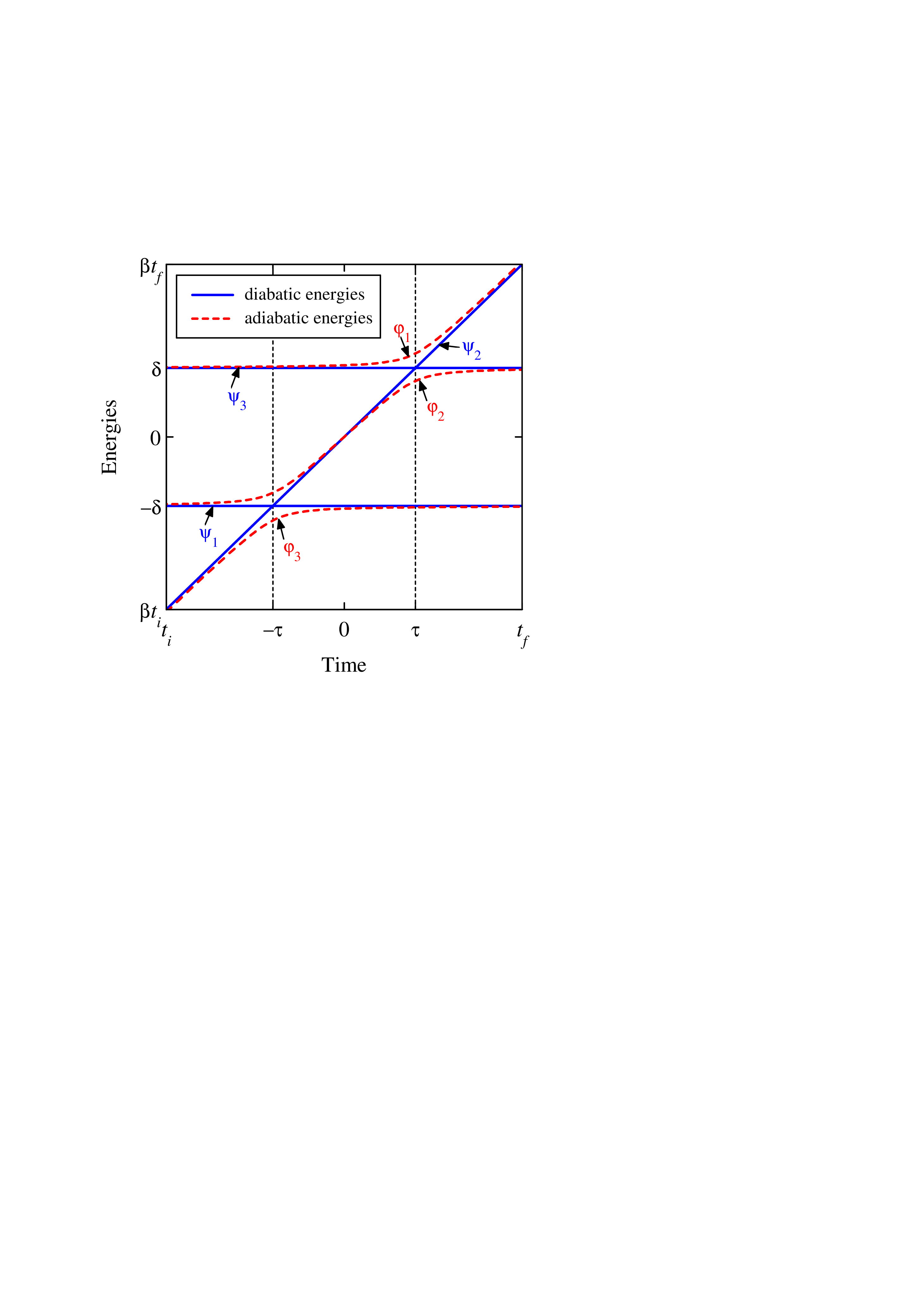}
\caption{(Color online) Diabatic and adiabatic energies in the three-state crossing model studied in this paper. 
We are concerned primarily with the (counterintuitive) transition $\protect\psi_{3}\rightarrow \protect\psi _{1}$.}
\label{Fig-system}
\end{figure}

A very interesting feature of all grid models is that \emph{counterintuitive
transitions}, for which the level crossings appear in a \textquotedblleft
wrong\textquotedblright\ order, are forbidden. In the three-state example in
Fig. \ref{Fig-system} such is the transition $\psi _{3}\rightarrow \psi _{1}$%
. In the adiabatic limit, the inhibition of such transitions is easily
understood: for the $\psi _{3}\rightarrow \psi _{1}$ transition an adiabatic
path requires the level crossing between $\psi _{3}$ and $\psi _{2}$ to
occur before the crossing between $\psi _{2}$ and $\psi _{1}$; this is not
the case here and hence there is no adiabatic path linking $\psi _{3}$ to $%
\psi _{1}$. In the general case of nonadiabatic evolution, however, the
inhibition of the $\psi _{3}\rightarrow \psi _{1}$ transition is not so
obvious because the concerned final state $\psi _{1}$ acquires some \emph{%
nonzero} transient population during the interaction. Yet in the end this
population vanishes, and this is an exact result. A similar conclusion
applies to more general level-crossing models \cite{Sinitsyn04,Volkov05} as
well.

We have verified with numerical simulations that all ingredients of the
multilevel LZ models are essential for this feature: linear energies,
constant interactions and infinite duration. Nonlinear energies, pulsed
interactions or finite interaction duration can each lead to nonzero
probability for counterintuitive transitions.

Yurovsky \emph{et al.} \cite{Yurovsky99} have studied analytically and
numerically the counterintuitive transition probability in two variations of
the DO model: (i) with a finite interaction duration and (ii) with a
piecewise-linear sloped potential. They have used a perturbative approach
assuming a quasidegenerate band of parallel energies and have found nonzero
probabilities for counterintuitive transitions in both models.

In this paper, we derive analytically the probability for a counterintuitive
transition in the simplest case of the DO model involving three states, with
two parallel and one slanted energy, as shown in Fig. \ref{Fig-system}. We
assume a \emph{finite} interaction duration, but our approach does not use
the quasidegeneracy assumption of Yurovsky \emph{et al.} \cite{Yurovsky99};
therefore our results are more general, as far as the three-state case is
concerned. The purpose of this work is not just to show that the probability for
a counterintuitive transition is nonzero but rather to derive accurate
analytical estimates for it. Our approach involves transformation to the
adiabatic basis where the evolution is represented as a sequence of
instantaneous two-state LZ transitions at each crossing and adiabatic
evolution elsewhere. This approach allows us to derive the transition
probabilities between each pair of diabatic states, including the
probability for the counterintuitive transition.

The problem of counterintuitive transitions, besides quite interesting by
itself, has interesting physical implications, which are discussed in some
detail in Sec. \ref{Sec-experiment}. Amongst them, we mention the problem of
saturated optical schielding with near-resonant light, which plays an important role in cold atomic
collisions \cite{Suominen96a,Suominen96b,Yurovsky97,Napolitano,Piilo}.

This paper is organized as follows. We define the problem in Sec. \ref%
{Sec-definition} and the propagator is derived in Sec. \ref{Sec-propagator}
in the general case. The transition probabilities in the finite DO model
are derived and compared with numerical results in Sec. \ref{Sec-finiteDO}.
The time-dependent probabilities in the original (infinite) DO model are
presented in Sec. \ref{Sec-DO}. Some physical examples of counterintuitive
transitions are discussed in Sec. \ref{Sec-experiment}. The conclusions are
summarized in Sec. \ref{Sec-conclusions}.


\section{Definition of the problem\label{Sec-definition}}

\subsection{The system}

The probability amplitudes of the three-state system $\mathbf{C}%
(t)=[C_{1}(t),C_{2}(t),C_{3}(t)]^{T}$ with the energies shown in Fig. \ref%
{Fig-system} satisfy the Schr\"{o}dinger equation ($\hbar =1$), 
\begin{equation}
i\mathbf{\dot{C}}(t)=\mathsf{H}(t)\mathbf{C}(t),  \label{SEq-c}
\end{equation}%
where the overdot denotes $d/dt$. The Hamiltonian in the usual rotating-wave
approximation is given by \cite{Shore}%
\begin{equation}
\mathsf{H}(t)=\left[ 
\begin{array}{ccc}
-\delta  & \Omega _{12} & 0 \\ 
\Omega _{12} & \beta t & \Omega _{23} \\ 
0 & \Omega _{23} & \delta 
\end{array}%
\right] .  \label{Hc}
\end{equation}%
As the Hamiltonian (\ref{Hc}) shows, there is no direct coupling between
states $\psi _{1}$ and $\psi _{3}$ but each of them is coupled to state $%
\psi _{2}$. 
Without loss of generality the constant couplings $\Omega _{12}$ and $\Omega _{23}$ will be assumed real and positive. 
Both couplings are supposed to have the same finite duration, being turned on at time $t_{i}$ and turned off at time $t_{f}$. 
In the original DO model the couplings last from $-\infty $ to $+\infty $. 
Furthermore, the energy splitting parameter $\delta$ and the slope $\beta $ of the energy of state $\psi _{2}$
 are assumed positive too, 
\begin{equation}
\delta >0,\qquad \beta >0.
  \label{>0}
\end{equation}

Given these assumptions, the crossing between the diabatic energies of
states $\psi _{1}$ and $\psi _{2}$, occuring at time $t_{-}=-\tau $ precedes
the crossing between states $\psi _{2}$ and $\psi _{3}$, occuring at time $%
t_{+}=\tau $, where 
\begin{equation}
\tau =\frac{\delta }{\beta }.  \label{tau}
\end{equation}%
Therefore, the transition $\psi _{1}\rightarrow \psi _{3}$ is \emph{intuitive%
}, while the opposite transition $\psi _{3}\rightarrow \psi _{1}$ is \emph{%
counterintuitive}. In the adiabatic limit, the transition probability from $%
\psi _{1}$ to $\psi _{3}$ is $P_{1\rightarrow 3}=1$, whereas that from $\psi
_{3}$ to $\psi _{1}$ is $P_{3\rightarrow 1}=0$.

\subsection{The Demkov-Osherov model}

In the DO model the transition probabilities $P_{m\rightarrow n}$ from state 
$\psi _{m}$ at $t\rightarrow -\infty $ to state $\psi _{n}$ at $t\rightarrow
+\infty $ are given exactly by products of two-state single-crossing LZ
probabilities, as follows%
\begin{equation}
\begin{array}{lll}
P_{1\rightarrow 1}=p_{-}, & P_{1\rightarrow 2}=q_{-}p_{+}, & P_{1\rightarrow
3}=q_{-}q_{+}, \\ 
P_{2\rightarrow 1}=q_{-}, & P_{2\rightarrow 2}=p_{-}p_{+}, & P_{2\rightarrow
3}=p_{-}q_{+}, \\ 
P_{3\rightarrow 1}=0, & P_{3\rightarrow 2}=q_{+}, & P_{3\rightarrow 3}=p_{+},%
\end{array}
\label{P-DO}
\end{equation}%
where%
\begin{equation}
p_{\pm }=e^{-2\pi \alpha _{\pm }^2},\qquad q_{\pm }=1-p_{\pm },  \label{p}
\end{equation}%
i.e. $q_{\pm }$ is the transition probability and $p_{\pm }$ is the
probability of no transition at the crossing $t_{\pm }$, with 
\begin{equation}
\alpha _{-}=\Omega _{12}/\beta ^{\frac{1}{2}},\qquad \alpha _{+}=\Omega_{23}/\beta ^{\frac{1}{2}}.  \label{alphas}
\end{equation}

These simple results coincide with what would be expected naively, by
treating the crossings independently, no matter how close they are to each
other, and multiplying LZ probabilities. In particular, if the system is
initially in state $\psi _{3}$, the transition probability to state $\psi
_{1}$ is exactly zero at $t\rightarrow +\infty $, $P_{3\rightarrow 1}=0$,
which means that the counterintuitive transition $\psi _{3}\rightarrow \psi
_{1}$ is forbidden. This zero probability is rather unexpected because state 
$\psi _{1}$ acquires some nonzero population during the interaction.
However, it vanishes at $t\rightarrow +\infty $ for \emph{any} set of
parameters, irrespective of whether the interaction is adiabatic or not.
This property is unique for the DO model and it depends crucially on any of
its features: infinite coupling durations, constant couplings, constant
energies of states $\psi _{1}$ and $\psi _{3}$ and linear energy of state $\psi _{2}$. 
The goal of the present paper is to estimate the probability for counterintuive transitions in the case of finite coupling duration. 


\section{Transition matrix\label{Sec-propagator}}

\subsection{Eigenvalues and eigenstates}


We need the eigenvalues and the eigenstates (the adiabatic states) of $\mathsf{H}(t)$. 
The eigenvalues read \cite{Shore} 
\begin{subequations}
\label{eigenvalues}
\begin{eqnarray}
\lambda _{1} &=&-\frac{1}{3}a+\frac{2}{3}s\cos \frac{1}{3}\theta ,
\label{EV1} \\
\lambda _{2} &=&-\frac{1}{3}a-\frac{2}{3}s\cos \frac{1}{3}(\theta +\pi ),
\label{EV2} \\
\lambda _{3} &=&-\frac{1}{3}a-\frac{2}{3}s\cos \frac{1}{3}(\theta -\pi ),
\label{EV3}
\end{eqnarray}%
where 
\end{subequations}
\begin{subequations}
\label{abc}
\begin{eqnarray}
a &=&-\beta t,  \label{a} \\
b &=&-(\delta^2+\Omega _{12}^2+\Omega _{23}^2),  \label{b} \\
c &=&\delta (\Omega _{12}^2-\Omega _{23}^2+\delta \beta t),  \label{c} \\
s &=&\sqrt{a^2-3b},  \label{s} \\
\cos \theta  &=&-\frac{2a^{3}-9ab+27c}{2s^{3}}.  \label{theta}
\end{eqnarray}%
The eigenstates are given by $\varphi _{k}=[f_{1k},f_{2k},f_{3k}]^{T}$, with 
\end{subequations}
\begin{subequations}
\label{ES}
\begin{eqnarray}
f_{1k} &=&\frac{1}{N_{k}}\Omega _{12}(\lambda _{k}-\delta ),  \label{ES1} \\
f_{2k} &=&\frac{1}{N_{k}}(\lambda _{k}^2-\delta^2),  \label{ES2} \\
f_{3k} &=&\frac{1}{N_{k}}\Omega _{23}(\lambda _{k}+\delta ),  \label{ES3}
\end{eqnarray}%
where $N_{k}$ are normalization factors ($k=1,2,3$). The asymptotic
behaviors at large times of the eigenvalues and the eigenstates are
presented in Appendix \ref{Sec-asymptotic}.


\subsection{Adiabatic basis}

The transformation linking the diabatic amplitudes $\mathbf{C}(t)$ and the
adiabatic amplitudes $\mathbf{A}(t)$ is given by 
\end{subequations}
\begin{equation}
\mathbf{C}(t)=\mathsf{F}(t)\mathbf{A}(t),  \label{transformation}
\end{equation}%
where $\mathbf{A}(t)=[A_{1}(t),A_{2}(t),A_{3}(t)]^{T}$ and $\mathsf{F}(t)$
is an orthogonal rotation matrix [$\mathsf{F}^{-1}(t)=\mathsf{F}^{T}(t)$]
whose columns are the eigenvectors (\ref{ES}), 
\begin{equation}
\mathsf{F}(t)=\left[ 
\begin{array}{ccc}
f_{11}(t) & f_{12}(t) & f_{13}(t) \\ 
f_{21}(t) & f_{22}(t) & f_{23}(t) \\ 
f_{31}(t) & f_{32}(t) & f_{33}(t)%
\end{array}%
\right] .  \label{F(t)}
\end{equation}%
The Schr\"{o}dinger equation in the adiabatic basis reads 
\begin{equation}
i\mathbf{\dot{A}}(t)=\mathsf{H}_{A}(t)\mathbf{A}(t),  \label{SEq-a}
\end{equation}%
with $\mathsf{H}_{A}(t)=\mathsf{F}^{T}(t)\mathsf{H}(t)\mathsf{F}(t)-i\mathsf{%
F}^{T}(t)\dot{\mathsf{F}}(t)$, or 
\begin{equation}
\mathsf{H}_{A}(t)=\left[ 
\begin{array}{ccc}
\lambda _{1} & -i\nu _{12} & -i\nu _{13} \\ 
-i\nu _{21} & \lambda _{2} & -i\nu _{23} \\ 
-i\nu _{31} & -i\nu _{32} & \lambda _{3}%
\end{array}%
\right] ,  \label{H-a}
\end{equation}%
where the nonadiabatic coupling between the adiabatic states $\varphi _{k}(t)$
and $\varphi _{l}(t)$ is
\begin{equation}
\nu _{kl}(t)=\langle \varphi _{k}(t)\mid \dot{\varphi}_{l}(t)\rangle =-\nu
_{lk}(t)  \label{NAC}
\end{equation}%

We use the fact that the transition times in the adiabatic
basis are shorter than in the diabatic basis \cite{LZtimes}. This is so
because while the asymptotic behaviors of the adiabatic energies at large
times (\ref{EV plus}) are approximately the same as the asymptotics of the
diabatic energies, the couplings $\nu _{kl}$ in the adiabatic
basis (\ref{NAC}) vanish as $t^{-2}$ [see Eqs. (\ref{EV plus})], in contrast
to the constant couplings $\Omega _{12}$ and $\Omega _{23}$ in the diabatic
basis. The difference in the transition times is illustrated in Fig. \ref%
{Fig-evolution}, where the oscillations in the populations of the adiabatic
states vanish much faster.

\begin{figure}[t]
\includegraphics[width=75mm]{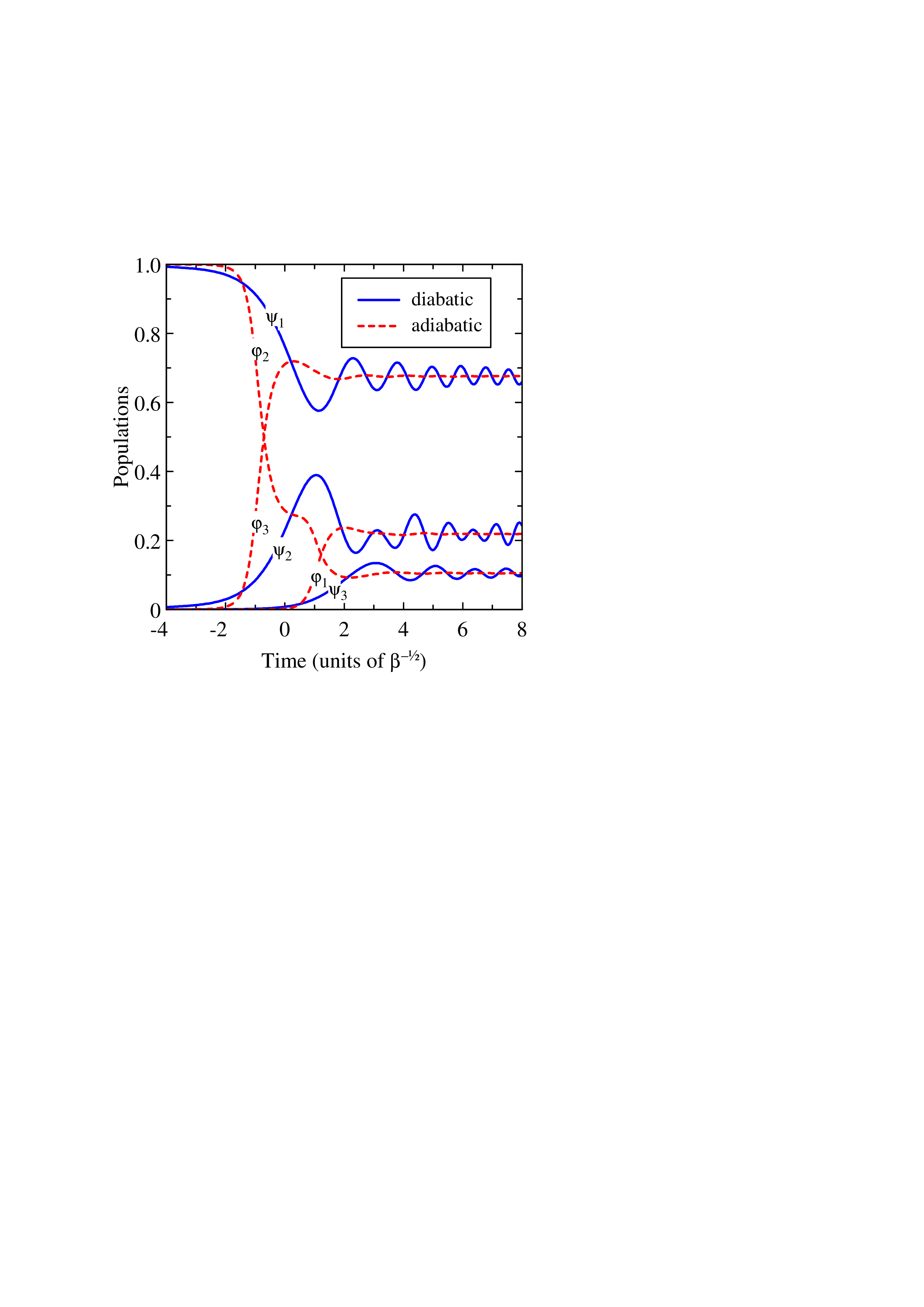}
\caption{(Color online) Time evolutions of the populations of the diabatic and adiabatic states for the original DO model ($t_i=-\infty$). 
The system starts in state $\protect\psi _{1}$. 
The interaction parameters are $\Omega _{12}=\Omega _{23}=\protect\beta ^{\frac{1}{2}}$,
 $\protect\delta =\protect\beta ^{\frac{1}{2}}$. }
\label{Fig-evolution}
\end{figure}


\subsection{Evolution matrix in the adiabatic basis}

Our method is based on two simplifying assumptions. 
First, we assume that appreciable transitions take place only between neighboring adiabatic
states, $\varphi _{1}(t)\leftrightarrow \varphi _{2}(t)$ and $\varphi_{2}(t)\leftrightarrow \varphi _{3}(t)$,
 but not between states $\varphi_{1}(t)$ and $\varphi _{3}(t)$, because the energies of the latter pair are
split by the largest gap. Second, we assume that the nonadiabatic
transitions occur instantly at the corresponding avoided crossings and the evolution is adiabatic elsewhere. 
This allows us to obtain the propagator in the adiabatic basis by multiplying five simple transition matrices
 describing LZ transitions or adiabatic evolution. 

The adiabatic evolution matrix $\mathsf{U}^{A}(t_{f},t_{i})$ is most
conveniently determined in the adiabatic interaction representation, where
the diagonal elements of $\mathsf{H}_{A}(t)$ are nullified. The
transformation reads 
\begin{equation}
\mathbf{A}(t)=\mathsf{M}(t)\mathbf{B}(t),  \label{a=Mb}
\end{equation}%
where 
\begin{equation}
\mathsf{M}(t,t_{0})=\left[ 
\begin{array}{ccc}
e^{-i\Lambda _{1}(t,t_{0})} & 0 & 0 \\ 
0 & e^{-i\Lambda _{2}(t,t_{0})} & 0 \\ 
0 & 0 & e^{-i\Lambda _{3}(t,t_{0})}%
\end{array}%
\right] ,  \label{M(t,t0)}
\end{equation}
\begin{subequations}
\label{Lambdas}
\begin{eqnarray}
\Lambda _{k}(t,t_{0}) &=&\int_{t_{0}}^{t}\lambda _{k}(t^{\prime })dt^{\prime
},  \label{Lambda} \\
\Lambda _{kl}(t,t_{0}) &\equiv &\Lambda _{k}(t,t_{0})-\Lambda _{l}(t,t_{0}),
\end{eqnarray}%
and $t_{0}$ is an arbitrary fixed time. The Schr\"{o}dinger equation in this basis reads
\end{subequations}
\begin{equation}
i\mathbf{\dot{B}}(t)=\mathsf{H}_{B}(t)\mathbf{B}(t),  \label{SEq-b}
\end{equation}%
with%
\begin{equation}
\mathsf{H}_{B}(t)=-i\left[ 
\begin{array}{ccc}
0 & \nu _{12}e^{i\Lambda _{12}(t,t_{0})} & \nu _{13}e^{i\Lambda
_{13}(t,t_{0})} \\ 
\nu _{21}e^{i\Lambda _{21}(t,t_{0})} & 0 & \nu _{23}e^{i\Lambda
_{23}(t,t_{0})} \\ 
\nu _{31}e^{i\Lambda _{31}(t,t_{0})} & \nu _{32}e^{i\Lambda _{32}(t,t_{0})}
& 0%
\end{array}%
\right] .  \label{Hb}
\end{equation}
In this basis, the evolution matrix for adiabatic evolution is given by the identity matrix.

The LZ transitions at the crossings at $\pm \tau $ are described by the transition matrices 
\begin{subequations}
\label{Ulz}
\begin{eqnarray}
\mathsf{U}_{LZ}(-\tau ) &=&\left[ 
\begin{array}{ccc}
1 & 0 & 0 \\ 
0 & \sqrt{q_{-}}e^{-i\phi _{-}} & -\sqrt{p_{-}} \\ 
0 & \sqrt{p_{-}} & \sqrt{q_{-}}e^{i\phi _{-}}%
\end{array}%
\right] , \\
\mathsf{U}_{LZ}(\tau ) &=&\left[ 
\begin{array}{ccc}
\sqrt{q_{+}}e^{-i\phi _{+}} & -\sqrt{p_{+}} & 0 \\ 
\sqrt{p_{+}} & \sqrt{q_{+}}e^{i\phi _{+}} & 0 \\ 
0 & 0 & 1%
\end{array}%
\right] ,
\end{eqnarray}%
\end{subequations}
where $p_{\pm }$ and $q_{\pm }$ are given by Eqs. (\ref{p}) and 
\begin{equation}
\phi _{\pm }=\arg \Gamma (1-i\alpha _{\pm }^2)+\frac{\pi }{4}+ \alpha_{\pm}^2
 \left( \ln \alpha _{\pm }^2-1\right) ,
\label{phi}
\end{equation}%
with $\alpha _{\pm }$ given by Eqs. (\ref{alphas}). 
The LZ phases $\phi _{\pm }$ do not depend on time, unlike the dynamical phases (\ref{Lambda}).

The propagator in the adiabatic basis reads $\mathsf{U}^{A}(t_{f},t_{i})=%
\mathsf{M}(t_{f},\tau )\mathsf{U}_{LZ}(\tau )\mathsf{M}(\tau ,-\tau )\mathsf{U}_{LZ}(-\tau )\mathsf{M}(-\tau ,t_{i})$, or 
\begin{widetext}
\be\label{UA}
{\sf U}^{A}(t_f,t_i) = \left[ \begin{array}{ccc}
\sqrt{q_{+}}e^{-i\LZphase_{+}-i\EVA_1(t_f,t_i)} &
 -\sqrt{p_{+}q_{-}}e^{-i\LZphase_{-}-i\EVA_1(t_f,\tau )-i\EVA_2(\tau ,t_i)}
 & \sqrt{p_{-}p_{+}}e^{-i\EVA_1(t_f,\tau)-i\EVA_2(\tau ,-\tau )-i\EVA_3(-\tau ,t_i)} \\ 
\sqrt{p_{+}}e^{-i\EVA_1(\tau ,t_i)-i\EVA_2(t_f,\tau)}
 & \sqrt{q_{-}q_{+}}e^{i(\LZphase_{+}-\LZphase_{-})-i\EVA_2(t_f,t_i)}
 & -\sqrt{p_{-}q_{+}}e^{i\LZphase_{+}-i\EVA_2(t_f,-\tau )-i\EVA_3(-\tau ,t_i)} \\ 
0 & \sqrt{p_{-}}e^{-i\EVA_2(-\tau ,t_i)-i\EVA_3(t_f,-\tau )}
 & \sqrt{q_{-}}e^{i\LZphase_{-}-i\EVA_3(t_f,t_i)}
\end{array} \right] ,
\ee 
In the special case when $t_i=-T$, $t_f=T$ and $\Omega _{12}=\Omega_{23}\equiv \Omega $,
 many expressions simplify, as shown in Appendix \ref{Sec-symmetric}. 
Then $\alpha _{+}=\alpha _{-}\equiv\alpha$, $p_{+}=p_{-}\equiv p$, $q_{+}=q_{-}\equiv q=1-p$, and Eq. (\ref{UA}) reduces to
\be\label{U-symmetric}
{\sf U}^{A}(T,-T)=\left[ \begin{array}{ccc}
\sqrt{q}\ e^{-i\LZphase-i\EVA_1(T,-T)} & -\sqrt{pq}\ e^{-i\LZphase-i\EVA_{12}(T,\tau )} & p \\ 
\sqrt{p}\ e^{i\EVA_3(T,-\tau )-i\EVA_2(T,\tau )} & q & -\sqrt{pq}\ e^{i\LZphase+i\EVA_{12}(T,\tau )} \\ 
0 & \sqrt{p}\ e^{i\EVA_2(T,\tau )-i\EVA_3(T,-\tau )} & \sqrt{q}\ e^{i\LZphase+i\EVA_1(T,-T)}
\end{array} \right] .
\ee
\end{widetext}


\subsection{Evolution matrix in the diabatic basis}

The propagator in the diabatic basis can be obtained by using the transformation (\ref{transformation}); it reads
\begin{equation}
\mathsf{U}(t_{f},t_{i})=\mathsf{F}(t_{f})\mathsf{U}^{A}(t_{f},t_{i})\mathsf{F}^{T}(t_{i}).  
\label{U}
\end{equation}%
We shall use this relation to derive the transition probabilities in the finite and original DO models below.


\section{Transition Probabilities in the finite Demkov-Osherov model\label%
{Sec-finiteDO}}

\subsection{The propagator}

In order to obtain simpler formulas for the probabilities we assume that $t_{i}=-T$, $t_{f}=T$,
 although our approach is not limited to these restrictions. 
The transition probability from state $\psi _{m}$ to $\psi _{n}$
 is given by $P_{m\rightarrow n}=\left\vert U_{nm}(T,-T)\right\vert^2$, where 
\begin{equation}
U_{nm}(T,-T)=\sum_{k,l=1}^{3}f_{nk}(T)U_{kl}^{A}(T,-T)f_{ml}(-T).
\label{Unm}
\end{equation}%
Using this relation one can calculate the transition probability between any two states of the system. 
We pay special attention to the probability for counterintuitive transitions $P_{3\rightarrow 1}$,
 which is zero in the original DO model.


\subsection{Counterintuitive transition}

In the special case of equal couplings, $\Omega_{12}=\Omega_{23}\equiv \Omega $,
 we find from Eqs. (\ref{U-symmetric}), (\ref{Unm}) and (\ref{F(-T)}) that
 the transition probability $P_{3\rightarrow 1}=|U_{13}(T,-T)|^2$ reads
\begin{eqnarray}
P_{3\rightarrow 1}&=& \big|pf_{11}^2+qf_{12}^2
 -2\sqrt{pq}f_{11}f_{12}\cos [\Lambda _{12}(T,\tau )+\phi] \notag \\
&& +2\sqrt{p}f_{12}f_{13}\cos [\Lambda _{2}(T,\tau )-\Lambda _{3}(T,-\tau )]  \notag \\
&& +2\sqrt{q}f_{11}f_{13}\cos[\Lambda_1(T,-T)+\phi]\big|^2.  
\label{P31}
\end{eqnarray}
It can be written as
\begin{equation}
P_{3\rightarrow 1} = \overline{P_{3\rightarrow 1}}+\widetilde{P_{3\rightarrow 1}},
  \label{P31sum}
\end{equation}
where $\overline{P_{3\rightarrow 1}}$ is the \emph{average probability}
 and $\widetilde{P_{3\rightarrow 1}}$ is the oscillating part.
By using the asymptotic expansions (\ref{ASasymptotics}) for $f_{mn}$ in Appendix \ref{Sec-asymptotic}
 and keeping the leading terms in the expansion over $1/T$ we find
\begin{subequations}
\label{P31avr-approximation}
\begin{eqnarray}\label{P31avr-exact}
\overline{P_{3\rightarrow 1}} &=& (pf_{11}^2+qf_{12}^2)^2+2(qf_{11}^2+pf_{12}^2)f_{13}^2+2pqf_{11}^2f_{12}^2 \\
&\sim & \frac{\Omega^2 (\Omega^2p+4\delta^2q)}{2\delta^2\beta^2T^2}
 + \frac{\Omega^2(\Omega^4p-8\delta^4q)}{2\delta^3 \beta^3 T^3} + \dots,
  \label{P31avr-largeT}
\end{eqnarray}%
\end{subequations}
\begin{eqnarray}
\widetilde{P_{3\rightarrow 1}} & \sim& \frac{\Omega^2}{\beta^2T^2}\left\{4q\cos^2[\Lambda_1(T,-T)+\phi]\right.\notag \\
&& +\frac{p\Omega^2}{\delta^2}\cos^2\left[ \Lambda _{2}(T,\tau)-\Lambda _{3}(T,-\tau )\right] \notag \\
&& -\frac{4\sqrt{pq}\Omega }{\delta } \cos [\Lambda _{1}(T,-T)+\phi]  \notag\\
&& \times \left. \cos \left[ \Lambda _{2}(T,\tau )-\Lambda _{3}(T,-\tau )%
\right] \right\}+\dots 
\end{eqnarray}

For unequal couplings ($\Omega_{12}\neq\Omega_{23}$), the expansion over $1/T$ of the average probability reads
\begin{eqnarray}
\overline{P_{3\rightarrow 1}} &\sim& 
 \frac{ \Omega_{12}^2\Omega_{23}^2(p_-+p_+) + 4\delta^2(\Omega_{12}^2q_++\Omega_{23}^2q_-)}
 {4\delta^2\beta^2T^2}\notag \\
&+& 
 \frac{ \Omega_{12}^2\Omega_{23}^2(\Omega_{12}^2p_-+\Omega_{23}^2p_+) - 8\delta^4(\Omega_{12}^2q_++\Omega_{23}^2q_-)}
 {4\delta^3\beta^3T^3}\notag \\
&+& \dots
\label{P31aver}
\end{eqnarray}%
The part $\widetilde{P_{3\rightarrow 1}}$ is too cumbersome to be presented here.

In the \emph{near-adiabatic regime} ($\alpha\gg 1$, $p\ll 1$, 
$q\approx 1$, and $\phi\ll 1$), we find from Eq. (\ref{P31}) that
\begin{equation}
P_{3\rightarrow 1}\sim \frac{4\Omega^2}{\beta^2T^2}\cos^2[\Lambda_{1}(T,-T)+\phi].  \label{P31-adiab}
\end{equation}

In the \emph{weak-coupling limit} ($\alpha\ll 1$, $p\approx 1$, $q\ll 1$, and $\phi\approx \pi/4 $),
 we find from Eq. (\ref{P31}) that 
\begin{equation}
P_{3\rightarrow 1}\sim \frac{\Omega ^{4}}{\delta^2\beta^2T^2}\cos
^2[\Lambda _{2}(T,\tau )-\Lambda _{3}(T,-\tau )].  \label{P31-weak}
\end{equation}

\begin{figure}[t]
\includegraphics[width=75mm]{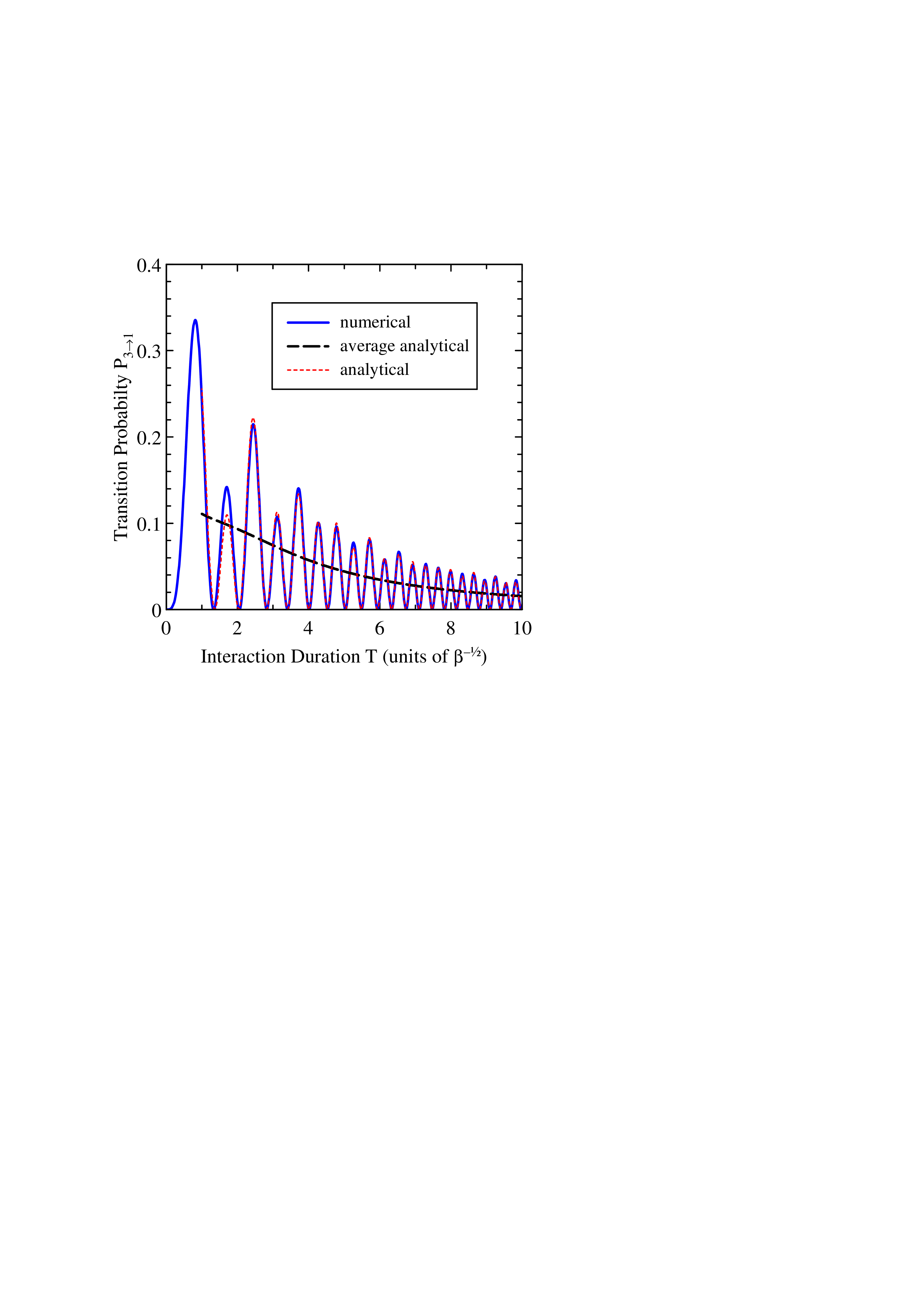}
\caption{(Color online)  Probability for counterintuitive transition $P_{3\rightarrow 1}$
plotted against the coupling duration $T$ for interaction parameters $\Omega
_{12}=\Omega _{23}=\protect\beta ^{\frac{1}{2}}$, $\protect\delta =\protect%
\beta ^{\frac{1}{2}}$. The full line shows numerical results, the thin dashed
curve the analytical approximation (\protect\ref{P31}) and the thick dashed
curve the average probability (\protect\ref{P31avr-approximation}).}
\label{Fig-T}
\end{figure}

Figure \ref{Fig-T} shows the counterintuitive transition probability $P_{3\rightarrow 1}$ against the coupling duration $T$. 
The probability decreases in an oscillatory manner, as predicted by our results.
The analytical approximation (\ref{P31}) describes very accurately both the phase and the amplitude of the oscillations.
The approximation (\ref{P31avr-approximation}) describes very accurately also the average
 probability $\overline{P_{3\rightarrow 1}}$.

Fig. \ref{Fig-delta} displays the counterintuitive transition probability $P_{3\rightarrow 1}$ 
 as a function of the energy separation parameter $\delta $. 
As with $T$, the probability decreases in an oscillatory manner. 
The analytical approximations are seen again to describe the probability very accurately.

\subsection{Other transition probabilities}

By using Eq. (\ref{Unm}) we can find all transition probabilities $%
P_{m\rightarrow n}=|U_{nm}(T,-T)|^2$ $(m,n=1,2,3)$. For the sake of
simplicity we assume again that $t_i=-T$, $t_f=T$ and $\Omega
_{12}=\Omega _{23}\equiv \Omega $, although our approach applies to the
general non-symmetric case as well. Using Eq. (\ref{U-symmetric}) we find
the average probabilities expanded to the lowest order of $1/T$,
\begin{subequations}\label{Pavr-finiteDO}
\begin{eqnarray}
\overline{P_{1\rightarrow 1}} &\sim &p+\frac{\Omega^2}{\beta^2T^2} [ \varkappa^2(q^2-2p)+1-2p-p^2] , \\
\overline{P_{1\rightarrow 2}} &\sim &pq+\frac{\Omega^2 }{\beta^2T^2}(\varkappa^2q^2+1-6p+7p^2), \\
\overline{P_{1\rightarrow 3}} &\sim &q^2+\frac{2\Omega^2}{\beta^2T^2}[\varkappa^2(p-q^2)+3pq-q] ,\\
\overline{P_{2\rightarrow 1}} &\sim &q+\frac{\Omega^2}{\beta^2T^2}(p^2+4p-3-\varkappa^2q^2) , \\
\overline{P_{2\rightarrow 2}} &\sim &p^2+\frac{2\Omega^2}{\beta^2T^2}(1+p-4p^2), \\
\overline{P_{2\rightarrow 3}} &\sim &pq+\frac{\Omega^2}{\beta^2T^2}(\varkappa^2q^2+1-6p+7p^2) , \\
\overline{P_{3\rightarrow 1}} &\sim &\frac{2\Omega^2}{\beta^2T^2}(\varkappa^2p+q),  \label{P31avr-T} \\
\overline{P_{3\rightarrow 2}} &\sim &q+\frac{\Omega^2}{\beta^2T^2}(p^2+4p-3-\varkappa^2q^2) , \\
\overline{P_{3\rightarrow 3}} &\sim &p+\frac{\Omega^2}{\beta^2T^2}[ \varkappa^2(q^2-2p)+1-2p-p^2] ,
\end{eqnarray}
where $\varkappa^2=\Omega^2/4\delta^2$. 
All these probabilities have the correct DO limits (\ref{P-DO}) for $T\rightarrow\infty$.
Note that
 $\overline{P_{1\rightarrow 1}}\sim \overline{P_{3\rightarrow 3}}$,
 $\overline{P_{1\rightarrow 2}}\sim \overline{P_{2\rightarrow 3}}$, and
 $\overline{P_{2\rightarrow 1}}\sim \overline{P_{3\rightarrow 2}}$.

\begin{figure}[t]
\includegraphics[width=75mm]{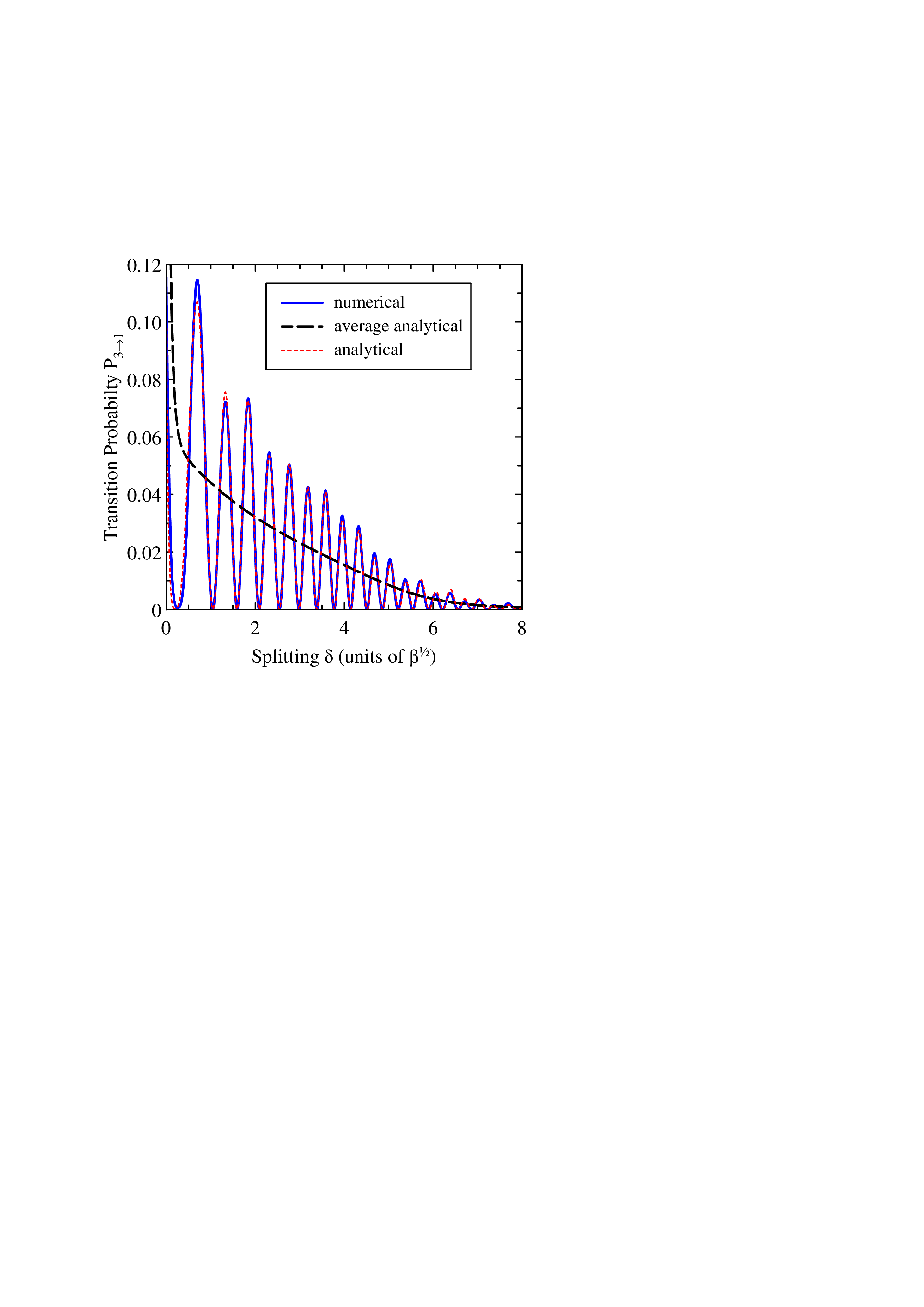}
\caption{(Color online)  Probability for counterintuitive transition $P_{3\rightarrow 1}$
plotted against the energy splitting parameter $\protect\delta $. 
The other interaction parameters
are $\Omega _{12}=\Omega _{23}=\protect\beta ^{\frac{1}{2}}$, $T=5\protect%
\beta ^{-\frac{1}{2}}$. The full line shows numerical results, the thin dashed
curve the analytical approximation (\protect\ref{P31}) and the thick dashed
curve the average probability (\protect\ref{P31avr-approximation}).}
\label{Fig-delta}
\end{figure}


\section{Time evolution in the original Demkov-Osherov model\label{Sec-DO}}

In the original DO model the time-dependent transition probability from
state $\psi _{m}$ to $\psi _{n}$ is given by $P_{m\rightarrow
n}(t)=\left\vert U_{nm}(t,-\infty )\right\vert^2$. 
We find from Eq. (\ref{U}) for $t_{i}=-\infty $, $t_{f}=t$ that 
\end{subequations}
\begin{equation}
U_{nm}(t,-\infty )=\sum_{k,l=1}^{3}f_{nk}(t)U_{kl}^{A}(t,-\infty
)f_{ml}(-\infty ),  \label{Unm-DO}
\end{equation}%
and we take into account that%
\begin{equation}
\mathsf{F}(-\infty )=\left[ 
\begin{array}{ccc}
0 & -1 & 0 \\ 
0 & 0 & 1 \\ 
1 & 0 & 0%
\end{array}%
\right] .  \label{F(-infty)}
\end{equation}


\subsection{Counterintuitive transition}

\begin{figure}[t]
\includegraphics[width=75mm]{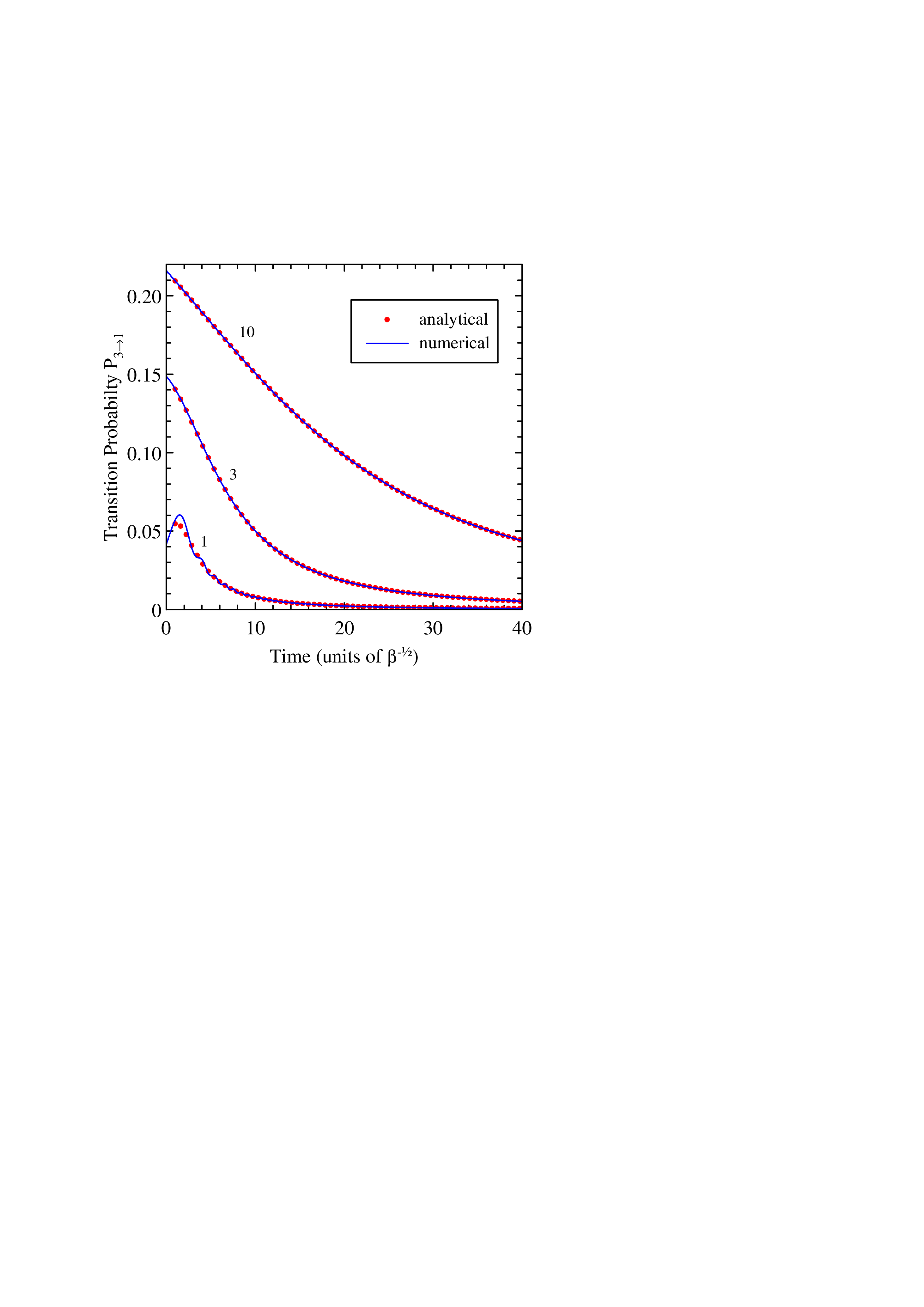}
\caption{(Color online)  Time evolution of the probability for counterintuitive transition
$P_{3\rightarrow 1}$ in the original DO model ($t_{i}=-\infty $) for $\protect\delta =\protect\beta ^{\frac12}$
 and three values of the couplings, $\Omega_{12}=\Omega_{23}=\protect\beta^{\frac12}$,
 $3\protect\beta^{\frac12}$, and $10\protect\beta^{\frac12}$ (denoted near the respective curves). 
The solid curves show numerical results and the dots show the analytical approximation
 (\protect\ref{P31avr-DO}) for the average probability. }
\label{Fig-time}
\end{figure}

By using Eq. (\ref{Unm-DO}) we find that the probability of counterintuitive
transition $P_{3\rightarrow 1}$ after the crossings reads
\begin{subequations}
\begin{eqnarray}
P_{3\rightarrow 1}(t) &=&|\sqrt{q_{+}}f_{11}(t)+\sqrt{p_{+}}%
f_{12}(t)e^{i[\Lambda _{12}(t,\tau )+\phi_+ ]}|^2 \\
&=&\overline{P_{3\rightarrow 1}(t)}+\widetilde{P_{3\rightarrow 1}(t)}
\end{eqnarray}%
where 
\end{subequations}
\begin{subequations}
\label{P31avr-DO}
\begin{align}
\overline{P_{3\rightarrow 1}(t)}& =q_{+}f_{11}^2+p_{+}f_{12}^2, \\
& \sim \frac{\Omega _{12}^2\left( 4\delta^2q_{+}+\Omega
_{23}^2p_{+}\right) }{4\delta^2\beta^2t^2},
\end{align}%
\end{subequations}
\begin{subequations}
\label{P31osc-DO}
\begin{align}
\widetilde{P_{3\rightarrow 1}(t)}& =2\sqrt{p_{+}q_{+}}f_{11}f_{12}\cos
[\Lambda _{12}(t,\tau )+\phi _{+}] \\
& \sim -\frac{\Omega _{12}^2\Omega _{23}}{\delta \beta^2t^2}\sqrt{%
p_{+}q_{+}}\cos [\Lambda _{12}(t,\tau )+\phi _{+}].  \label{P31osc-DO-approx}
\end{align}

Figure \ref{Fig-time} shows the time evolution of the probability for
counterintuitive transition $P_{3\rightarrow 1}$ for three values of the
couplings $\Omega _{12}=\Omega _{23}\equiv \Omega $. For these values
of $\Omega $ there are almost no oscillations visible, because $p_{+}\ll 1$
in Eq. (\ref{P31osc-DO-approx}). The probability $P_{3\rightarrow 1}(t)$
decreases with time, as predicted by Eq. (\ref{P31avr-DO}), and vanishes for
large times towards the DO result $P_{3\rightarrow 1}(\infty )=0$. 
As predicted, the transition probability $P_{3\rightarrow 1}(t)$ increases with the couplings.
The analytical approximation (\ref{P31avr-DO}) describes very accurately the average probability $\overline{P_{3\rightarrow 1}(t)}$.

\subsection{Other transition probabilities}

By using Eq. (\ref{Unm-DO}) and the large-$t$ asymptotic expansions in Appendix \ref{Sec-asymptotic},
 we find the leading terms of all average transition probabilities
 $\overline{P_{m\rightarrow n}(t)}$ after the crossings
 for equal couplings ($\Omega_{12}=\Omega _{23}\equiv \Omega$),
\end{subequations}
\begin{subequations}
\label{Pavr(t)}
\begin{eqnarray}
\overline{P_{1\rightarrow 1}(t)} &\sim &p+\frac{\Omega^2}{\beta^2t^2}%
\left[ \varkappa^2(q^2-p)-p^2\right] , \\
\overline{P_{1\rightarrow 2}(t)} &\sim &pq+\frac{\Omega^2}{\beta^2t^2}(1-3pq), \\
\overline{P_{1\rightarrow 3}(t)} &\sim &q^2+\frac{\Omega^2}{\beta
^2t^2}\left[ q\left( p-q\right) +\varkappa^2(p-q^2)\right] , \\
\overline{P_{2\rightarrow 1}(t)} &\sim &q+\frac{\Omega^2}{\beta^2t^2}%
\left( p^2-q-\varkappa^2q^2\right) , \\
\overline{P_{2\rightarrow 2}(t)} &\sim &p^2+\frac{\Omega^2}{\beta
^2t^2}(1-3p^2), \\
\overline{P_{2\rightarrow 3}(t)} &\sim &pq+\frac{\Omega^2}{\beta^2t^2%
}\left[ p(p-q)+\varkappa^2q^2\right] , \\
\overline{P_{3\rightarrow 1}(t)} &\sim &\frac{\Omega^2}{\beta^2t^2}%
\left( \varkappa^2p+q\right) ,  \label{P31avr(t)} \\
\overline{P_{3\rightarrow 2}(t)} &\sim &q+\frac{\Omega^2}{\beta^2t^2}%
(p-2q), \\
\overline{P_{3\rightarrow 3}(t)} &\sim &p+\frac{\Omega^2}{\beta^2t^2}%
\left( q-p-\varkappa^2p\right) ,
\end{eqnarray}%
\end{subequations}
where, as before, $\varkappa^2=\Omega^2/4\delta^2$. 
All these probabilities have the correct DO limits (\ref{P-DO}) for $t\rightarrow\infty$.
It is interesting to note that the counterintuitive-transition probability in the
original DO model (\ref{P31avr(t)}) is one-half of the one in the finite DO
model (\ref{P31avr-T}).

\section{Experimental implementations\label{Sec-experiment}}

This three-state DO model discussed here can be realized in several physical
systems. One example is when a chirped laser pulse couples an initially
populated state to a manifold of two levels simultaneously \cite{ARPC}, the
energy diagram for which is identical to the one in Fig. \ref{Fig-system},
with state $\psi _{2}$ being the initial state. Adiabatic evolution can
produce in this system a very selective excitation even if the Fourier
bandwidth of the laser pulse is larger than the level spacing within the manifold. 
Indeed, for \emph{red-to-blue chirp} ($\beta>0$) the system will follow from
left to right the lowest adiabatic energy, which links the initial state $%
\psi _{2}$ to the lowest diabatic (unperturbed) state $\psi _{1}$ of the manifold. 
In contrast, for \emph{blue-to-red chirp} ($\beta<0$) the system would follow (from right to left in Fig. \ref{Fig-system})
 the highest adiabatic energy, linking state $\psi_2$ to the highest diabatic state $\psi_3$. 
Therefore, the chirp sign alone determines if the population is directed towards the lowest or the
highest state of the manifold. 
This excitation scheme has been demonstrated experimentally by Warren and co-workers \cite{Melinger} on the $3s$-$3p$
transition in sodium. Red-to-blue chirped picosecond pulses populated
predominantly the lower fine-structure level $3p$ $^2P_{1/2}$, while
blue-to-red chirped pulses placed the population onto the upper
fine-structure level $3p$ $^2P_{3/2}$. Counterintuitive transitions can be
demonstrated in this system by using the same setup as described above, but
starting in one of the fine-structure levels.

Another example of the model discussed here is Stark-chirped rapid adiabatic
passage (SCRAP) \cite{SCRAP-theory,SCRAP-experiment,SCRAP-3S} where level
crossings are created by inducing ac Stark shifts of the energy levels by a
strong off-resonant laser pulse. A level diagram similar to the one in Fig. %
\ref{Fig-system} is found in the three-state version of SCRAP \cite{SCRAP-3S}.

It is also worth noting a similarity between the finite DO model presented here
and multiple-ground-state models of optical shielding of cold collisions in
magneto-optical atom traps \cite{Suominen96a,Suominen96b,Yurovsky97,Napolitano,Piilo}. 
Optical shielding techniques were originally developed to prevent laser cooled atoms from
escaping the trap by making the collisions between the atoms elastic \cite{Suominen96b}. 
The experiments show that the shielding process saturates, in contrast to
the expected complete shielding, when the intensity of the shielding field is
increased \cite{Suominen96b}. The reason for the saturation is still an open
problem and may have a connection to the counterintuitive transitions in the
corresponding multiple-ground-state level scheme of the quasimolecule, which
in its simple form resembles the present model.

\section{Conclusions\label{Sec-conclusions}}

We have calculated analytically the probability for the counterintuitive transition $\psi _{3}\rightarrow \psi _{1}$
 in a three-state system with crossing energies. 
We have performed the derivation in the adiabatic basis by assuming instantaneous
 transitions at the level crossings and adiabatic evolution elsewhere. 
The counterintuitive transition probability $P_{3\rightarrow 1}$ is nonzero at finite times,
 whereas it vanishes for infinite coupling duration, thus recovering the result in the DO model. 
A very good agreement with the numerical results is found. 
This approach has been used to derive the other transition probabilities within the three-state system too.

Our results suggest that at large $T$ the probability for a counterintuitive
transition is proportional to the factor $\Omega^2/(\delta^2\beta^2T^2)$. 
The decrease with the interaction duration $T$ is expected as for $T\rightarrow\infty$ the probability should vanish (DO model). 
The increase with the coupling $\Omega$ is also anticipated since larger interaction is expected to increase weak transitions. 
The decrease vs the energy splitting $\delta$ between the concerned states $\psi_1$ and $\psi_3$ is natural too. 
The decrease vs the slope $\beta$ is more subtle. 
Since $\beta T$ is the diabatic energy of state $\psi_2$,
 and since states $\psi_1$ and $\psi_3$ interact with each other via $\psi_2$,
 the effective coupling duration for the transition $\psi _{3}\rightarrow \psi _{1}$ depends on the slope $\beta$:
 the larger the slope the smaller the duration and therefore the smaller the transition probability.
The other (``intuitive'') transition probabilities exhibit similar dependences
 of the finite-duration corrections on the interaction parameters.

We have described physical examples of coherent atomic excitation where the
 described counterintuitive transition can be observed. 
The model studied here resembles also the simple form of the multiple-ground-state
quasimolecule model used to study optical shielding in magneto-optical atom
traps \cite{Suominen96a,Suominen96b}. The saturation of optical shielding in
high laser intensities has been earlier attributed to off-resonant
excitation to attractive states, counterintuitive transitions or other
processes including multiple partial waves \cite{Suominen96a,Suominen96b,Yurovsky97}. 
The results here show that the probability for counterintuitive transitions is clearly non-negligible
 and indicate that they can also play a role in the saturation of optical shielding.


\subsection*{Acknowledgements}


This work has been supported by the European Union's Transfer of Knowledge
project CAMEL (Grant No. MTKD-CT-2004-014427), and the Alexander von
Humboldt Foundation. AAR acknowledges support from the EU Marie Curie
Training Site project HPMT-CT-2001-00294.
JP acknowledges support from the Magnus Ehrnrooth Foundation
 and thanks for the hospitality during his visit to Sofia University.


\appendix


\section{Asymptotics of the eigenvalues and the eigenstates\label{Sec-asymptotic}}

Here we present the asymptotics of the eigenvalues (\ref{eigenvalues}) at
large positive time $t>0$, i.e. for $\beta t\gg \delta $ and $\beta t\gg
\Omega $. Then $a=-\beta t$, $b=-(\delta^2+\Omega _{12}^2+\Omega
_{23}^2)$, $c=\delta^2\beta t\left[ 1+\left( \Omega _{12}^2-\Omega
_{23}^2\right) /(\delta \beta t)\right] $, and
\begin{subequations}
\label{abc-asymptotic}
\begin{eqnarray}
s &\sim &\beta t+\frac{3\left( \delta^2+\Omega _{12}^2+\Omega
_{23}^2\right) }{2\beta t}, \\
\cos \theta  &\sim &1-\frac{27\delta^2}{2\beta^2t^2}, \\
\theta  &\sim &\frac{3\sqrt{3}\delta }{\beta t}.
\end{eqnarray}%
The eigenvalues have the asymptotics 
\end{subequations}
\begin{subequations}
\label{EV plus}
\begin{eqnarray}
\lambda _{1} &\sim &\beta t+\frac{\Omega _{12}^2+\Omega _{23}^2}{\beta t},
\\
\lambda _{2} &\sim &\delta -\frac{\Omega _{23}^2}{\beta t}, \\
\lambda _{3} &\sim &-\delta -\frac{\Omega _{12}^2}{\beta t}.
\end{eqnarray}%
\end{subequations}
Hence we find from Eqs. (\ref{ES}) that 
\begin{subequations}\label{ASasymptotics}
\begin{eqnarray}
\varphi _{1} &\sim& \left[ 
\frac{\Omega _{12}}{\beta t} ,\
1-\frac{\Omega _{12}^2+\Omega _{23}^2}{2\beta^2t^2} ,\
\frac{\Omega _{23}}{\beta t} 
\right]^T ,  \label{AS1}\\
\varphi _{2} &\sim& \left[ 
-\frac{\Omega _{12}\Omega _{23}}{2\delta \beta t},\
-\frac{\Omega _{23}}{\beta t},\
1-\frac{\Omega _{23}^2(\Omega _{12}^2+4\delta^2)}{8\delta^2\beta^2t^2}
\right]^T ,  \label{AS2} \\
\varphi _{3} &\sim& \left[ 
-1+\frac{\Omega_{12}^2(\Omega _{23}^2+4\delta^2)}{8\delta^2\beta^2t^2},\
\frac{\Omega _{12}}{\beta t},\ 
 -\frac{\Omega _{12}\Omega _{23}}{2\delta \beta t}
\right]^T .  \label{AS3}
\end{eqnarray}
\end{subequations}

\section{Symmetries of the eigenvalues and the eigenstates\label%
{Sec-symmetric}}

The eigenvalues and the eigenstates simplify when $t_i=-T$, $t_f=T$ and $\Omega _{12}=\Omega _{23}\equiv \Omega $. 
Then $\alpha _{+}=\alpha _{-}\equiv\alpha $, $p_{+}=p_{-}\equiv p$, $q_{+}=q_{-}\equiv q=1-p$,
 $a(T)=-\beta T$, $b(T)=-\delta^2-2\Omega^2$, $c(T)=\delta^2\beta T$, and 
\begin{subequations}
\label{parameters(-T)}
\begin{eqnarray}
s(T) &=&\sqrt{\beta^2T^2+3\delta^2+6\Omega^2}, \\
\cos \theta (T) &=&\frac{\beta T}{s^{3}}(\beta^2T^2-9\delta
^2+9\Omega^2).
\end{eqnarray}%
Hence $\theta (-T)=\pi -\theta (T)$ and therefore $\lambda _{1}(-T)=-\lambda
_{3}(T)$, $\lambda _{2}(-T)=-\lambda _{2}(T)$, $\lambda _{3}(-T)=-\lambda
_{1}(T)$, and 
\end{subequations}
\begin{subequations}
\label{EV symmetries}
\begin{eqnarray}
\Lambda _{2}(-\tau ,-T) &=&\Lambda _{2}(\tau ,T)=-\Lambda _{2}(T,\tau ), \\
\Lambda _{2}(T,-T) &=&\Lambda _{2}(\tau ,-\tau )=0, \\
\Lambda _{1}(-\tau ,-T) &=&\Lambda _{3}(\tau ,T)=-\Lambda _{3}(T,\tau ), \\
\Lambda _{3}(-\tau ,-T) &=&\Lambda _{1}(\tau ,T)=-\Lambda _{1}(T,\tau ).
\end{eqnarray}%
The transformation matrix at $-T$ is given by 
\end{subequations}
\begin{equation}\label{F(-T)}
\mathsf{F}(-T)=\left[ 
\begin{array}{ccc}
-f_{33}(T) & -f_{32}(T) & -f_{31}(T) \\ 
f_{23}(T) & f_{22}(T) & f_{21}(T) \\ 
-f_{13}(T) & -f_{12}(T) & -f_{11}(T)%
\end{array}%
\right] .
\end{equation}%
With these relations taken into account the propagator (\ref{UA}) reduces to Eq. (\ref{U-symmetric}).


\end{document}